\documentclass[smallabstract,smallcaptions]{dccpaper}

\usepackage[utf8]{inputenc}
\usepackage[T1]{fontenc}

\usepackage{amsmath}
\usepackage{amssymb}
\usepackage{graphicx}
\usepackage{textgreek}
\usepackage{url}

\newlength{\figurewidth}
\newlength{\smallfigurewidth}
\newcommand{\set}[1]{\ensuremath{\{ #1 \}}}
\newcommand{\abs}[1]{\ensuremath{\lvert #1 \rvert}}
\newcommand{\Oh}{\ensuremath{\mathsf{O}}}

\newcommand{\SA}{\textsf{SA}}

\newcommand{\BWT}{\textsf{BWT}}
\newcommand{\FMI}{\textsf{FMI}}

\newcommand{\RA}{\textsf{RA}}
\newcommand{\mSA}{\ensuremath{\mathsf{SA}}}

\newcommand{\mBWT}{\ensuremath{\mathsf{BWT}}}
\newcommand{\mC}{\ensuremath{\mathsf{C}}}
\newcommand{\mRA}{\ensuremath{\mathsf{RA}}}

\newcommand{\LF}{\textsf{LF}}
\newcommand{\rank}{\textsf{rank}}
\newcommand{\select}{\textsf{select}}
\newcommand{\mLF}{\ensuremath{\mathsf{LF}}}
\newcommand{\mrank}{\ensuremath{\mathsf{rank}}}
\newcommand{\mselect}{\ensuremath{\mathsf{select}}}

\newcommand{\Acoll}{\ensuremath{\mathcal{A}}}
\newcommand{\Bcoll}{\ensuremath{\mathcal{B}}}

\newcommand{\BWTmerge}{\textsf{BWT\nobreakdash-merge}}
\newcommand{\ropebwt}{\textsf{RopeBWT}}
\newcommand{\ropebwtii}{\textsf{RopeBWT2}}
\newcommand{\CEU}{\textsf{CEU}}
\newcommand{\RS}{\textsf{RS}}

\begin{document}

\title
{\vspace{-12pt}\large
\textbf{Burrows-Wheeler transform for terabases}
}

\author{%
Jouni Sirén\thanks{This work was supported by the Wellcome Trust grant [098051].} \\[0.5em]
{\small\begin{minipage}{\linewidth}\begin{center}
\begin{tabular}{c}
Wellcome Trust Sanger Institute \\
Wellcome Genome Campus \\
Hinxton, Cambridge, CB10 1SA, UK \\
\url{jouni.siren@iki.fi}
\end{tabular}
\end{center}\end{minipage}}
}

\maketitle
\thispagestyle{empty}

\begin{abstract}
In order to avoid the reference bias introduced by mapping reads to a reference genome, bioinformaticians are investigating reference-free methods for analyzing sequenced genomes. With large projects sequencing thousands of individuals, this raises the need for tools capable of handling terabases of sequence data. A key method is the Burrows-Wheeler transform (BWT), which is widely used for compressing and indexing reads. We propose a practical algorithm for building the BWT of a large read collection by merging the BWTs of subcollections. With our 2.4 Tbp datasets, the algorithm can merge 600 Gbp/day on a single system, using 30 gigabytes of memory overhead on top of the run-length encoded BWTs.
\end{abstract}

\Section{Introduction}

The decrease in the cost of DNA sequencing has flooded the world with \emph{sequence data}. The \emph{1000 Genomes Project} \cite{1000GP2015} sequenced the genomes of over 2500 humans, and there are other projects that are similar or greater in scale. A sequencing machine produces a large number of \emph{reads} (short sequences) that cover the genome many times over. For a 3~Gbp human genome, the total length of the reads is often 100~Gbp or more.

\emph{De novo assembly} of sequenced genomes is still too difficult to be routinely done. As a practical alternative, bioinformaticians usually align the reads to a \emph{reference genome} of the same species. Because most reference genomes come from the genomes of a small number of individuals, this introduces \emph{reference bias}, which may adversely affect the results of subsequent analysis. Switching from reference sequences to \emph{reference graphs} can reduce the bias, but such transition will likely take years \cite{Church2015}.

Preprocessing large datasets can take weeks. It is often not feasible to rebuild everything when new methods of analysis require new functionalities. Structures based on the \emph{Burrows-Wheeler transform} (\BWT) are often useful due to their versatility. A \emph{run-length encoded} \BWT{} compresses repetitive sequence collections quite well \cite{Maekinen2010}, while the similarities to the suffix tree and the suffix array make \BWT-based indexes suitable for many \emph{pattern matching} and \emph{sequence analysis} tasks \cite{Ohlebusch2013,Maekinen2015}.

The \emph{Read Server} project at the Sanger Institute develops tools for large-scale \emph{reference-free} genome analysis, avoiding reference bias. Unique reads are compressed and indexed using the \BWT, while metadata databases contain information on the original reads. Initially, the project works with
the low-coverage and exome data from phase 3 of the 1000 Genomes Project.
After error correction and trimming the reads to either 73~bp or 100~bp, the 922~billion original reads (86~Tbp) are reduced to 53.0~billion unique sequences (4.88~Tbp). These sequences are stored in 16 \BWT-based indexes \cite{Ferragina2005a} taking a total of 561.5~gigabytes.

The unique reads are partitioned between the \BWT{}s by the last two bases. Every query must be repeated in all 16 indexes. The \BWT{}s also require more space, as we cannot compress the similarities between the reads in different indexes. Reducing the number of indexes would improve both memory usage and query performance. This requires \BWT{} construction algorithms that can handle terabases of data.

There are four often contradictory requirements for large-scale \BWT{} construction:
\textbf{Speed.} Larger datasets require faster algorithms. As a rough guideline, an algorithm processing 1~Mbp/s is good for up to 100~Gbp, while remaining somewhat useful until 1~Tbp of data.
\textbf{Memory.} We may have to process $n$~bp datasets on systems with less than $n$~bits of memory.
\textbf{Hardware.} A single node in a typical computer cluster has tens of CPU cores, from tens to hundreds of gigabytes of memory, a limited amount of local disk space, and access to shared disk space with no performance guarantees. Algorithms using a GPU or a large amount of fast disk space require special-purpose hardware.
\textbf{Efficiency.} Large \BWT{}s can be built by doing a lot of redundant work on multiple nodes. As most computer clusters do not have large amounts of unused capacity, such inefficient algorithms are not suitable for repeated use.

The most straightforward approach to \BWT{} construction is to build a \emph{suffix array} using a fast general-purpose algorithm \cite{Mori2008,Nong2011}, and then derive the \BWT{} from the suffix array. These algorithms cannot be used with large datasets, as they require much more memory than the sequences themselves. Suffix arrays can be built on disk \cite{Gonnet1992}, but even the fastest algorithms cannot index the data faster than 1\nobreakdash--2~Mbp/s
\cite{Bingmann2013,Kaerkkaeinen2014a,Nong2014,Nong2015,Kaerkkaeinen2015a,Liu2015a}.

There are many \emph{direct} \BWT{} construction algorithms that do not need the suffix array. Some require a limited amount of working space on top of the \BWT{} \cite{Hon2007,Kaerkkaeinen2007,Siren2009,Okanohara2009}, while others use the disk as additional working space \cite{Ferragina2012,Beller2013}. These general-purpose algorithms rarely exceed 1\nobreakdash--2~Mbp/s. \emph{Specialized algorithms} for DNA sequences achieve better time/space trade-offs. Some can index 5\nobreakdash--10~Mbp/s using ordinary hardware, with their memory usage becoming the bottleneck after about 1~Tbp \cite{Bauer2013,Li2014a}. GPU-based algorithms are even faster, but their memory usage is also higher \cite{Liu2014,Pantaleoni2014}. Distributing the \BWT{} construction to multiple nodes can remove the obvious bottlenecks, at the price of using more resources for the construction \cite{Wang2015}.

In this paper, we propose a practical algorithm for building the \BWT{} for terabases of sequence data. The algorithm is based on dividing the sequence collection into a number of subcollections, building the \BWT{} for each subcollection, and \emph{merging} the \BWT{}s into a single structure \cite{Siren2009}. The merging algorithm is faster than \BWT{} construction for the subcollections, while having a relatively small memory overhead on top of the final \BWT-based index. As the index must be loaded in memory for use, it can be built on the same system as it is going to be used.

\Section{Background}

A \emph{string} $S[1,n] = s_{1} \dotsm s_{n}$ is a sequence of \emph{characters} over an \emph{alphabet} $\Sigma = \set{1, \dotsc, \sigma}$. For indexing purposes, we consider \emph{text} strings $T[1,n]$ terminated by an endmarker $T[n] = \$ = 0$ not occurring elsewhere in the text. \emph{Binary} sequences are strings over the alphabet $\set{0, 1}$. A \emph{substring} of string $S$ is a sequence of the form $S[i,j] = s_{i} \dotsm s_{j}$. We call substrings of the type $S[1,j]$ and $S[i,n]$ \emph{prefixes} and \emph{suffixes}, respectively.

The \emph{suffix array} (\SA) \cite{Manber1993} is a simple full-text index. Given a text $T$, its suffix array $\mSA_{T}[1,n]$ is an array of pointers to the suffixes of the text in \emph{lexicographic order}.\footnote{If the text is evident from the context, we will omit the subscript and write just \SA{}, \BWT{}, etc.} We can build the suffix array in $\Oh(n)$ time using $2n$ bits of working space on top of the text and the suffix array \cite{Nong2011}. Given a \emph{pattern} $P$, we can find the \emph{lexicographic range} $[sp,ep]$ of suffixes prefixed by the pattern in $\Oh(\abs{P} \log n)$ time. The range of pointers $\mSA[sp,ep]$ lists the \emph{occurrences} of the pattern in the text.

The suffix array requires several times more memory than the original text. For large texts, this can be a serious drawback. We can use the \emph{Burrows-Wheeler transform} (\BWT) \cite{Burrows1994} as a more space-efficient alternative to the suffix array. The \BWT{} is an easily reversible permutation of the text with a similar combinatorial structure to the suffix array. Given a text $T[1,n]$ and its suffix array, we can easily produce the \BWT{} as $\mBWT[i] = T[\mSA[i]-1]$ (with $\mBWT[i] = T[n]$, if $\mSA[i] = 1$).

If $X \le Y$ in lexicographic order, we also have $cX \le cY$ for any character $c$. If $\mBWT[i] = c$ is the $j$\nobreakdash-th occurrence of $c$ in the \BWT{} and $\mSA[i]$ points to suffix $X$, suffix $cX$ is the $j$\nobreakdash-th suffix starting with $c$ in lexicographic order.

Let $\mC[c]$ bet the number of suffixes starting with a character smaller than $c$, and let $S.\mrank(i,c)$ be the number of occurrences of $c$ in the prefix $S[1,i]$. We define \LF-\emph{mapping} as $\mLF(i,c) = \mC[c] + \mBWT.\mrank(i, c)$ and $\mLF(i) = \mLF(i,\mBWT[i])$. The general form $\mLF(i,c)$ is the number of suffixes $X$ of text $T$ with $X \le cT[\mSA[i],n]$. This is known as the  \emph{lexicographic rank} $\mrank(cT[\mSA[i],n],T)$ of text $cT[\mSA[i],n]$ among the suffixes of text $T$. The specific form $\mLF(i)$ gives the lexicographic rank of the previous suffix ($\mSA[\mLF(i)] = \mSA[i]-1$, if $\mSA[i] > 1$, and $\mSA[\mLF(i)] = n$ otherwise).

The \emph{FM-index} (\FMI) \cite{Ferragina2005a} is a full-text index based on the \BWT{}. We use \emph{backward searching} in the FM-index to find the lexicographic range $[sp,ep]$ matching pattern $P$. Let $[sp_{i},ep_{i}]$ be the range of suffixes of text $T$ matching suffix $P[i, \abs{P}]$ of the pattern. We find $[sp_{i-1},ep_{i-1}]$ as $[\mLF(sp_{i}-1, P[i-1]) + 1, \mLF(ep_{i}, P[i-1])]$. By starting from $[sp_{\abs{P}}, ep_{\abs{P}}] = [\mC[P[\abs{P}]]+1, \mC[P[\abs{P}]+1]]$, we can find the lexicographic range of suffixes starting with the pattern in $\Oh(\abs{P} \cdot t_{r})$ time, where $t_{r}$ is the time required to answer \rank{} queries on the \BWT. In practice, the time complexity ranges from $\Oh(\abs{P})$ to $\Oh(\abs{P} \log n)$, depending on the encoding of the \BWT{}.

The FM-index \emph{samples} some suffix array pointers, including the one to the beginning of the text. When unsampled pointers are needed, they are derived by using \LF\nobreakdash-mapping. If $\mSA[i]$ is not sampled, the FM-index proceeds to $\mLF(i)$ and continues from there. If $\mSA[\mLF^{k}(i)]$ is the first sample encountered, $\mSA[i] = \mSA[\mLF^{k}(i)] + k$.
Depending on the way the samples are selected, we may need a binary sequence to mark the pointers that have been sampled.

Assume that we have an ordered \emph{collection} of texts $\Acoll = (T_{1}, \dotsc, T_{m})$ of total length $n = \abs{\Acoll} = \sum_{i} \abs{T_{i}}$. We want to build a (generalized) \BWT{} for the collection. The usual way is to make all endmarkers distinct, giving the one at the end of text $T_{i}$ character value $(0,i)$. This also makes all suffixes of the collection distinct. To save space, we still encode each endmarker as a $0$ in the \BWT{}. Because of this, \LF\nobreakdash-mapping does not work with $c = 0$, and we cannot match patterns spanning text boundaries.

When the texts are short (e.g.~reads), there are more space-efficient alternatives to sampling. Because all endmarkers have distinct values during sorting, we know that $\mSA[i]$ with $i \le m$ points to the end of text $T_{i}$. To find the end, we iterate $\Psi(i) = \mBWT.\mselect(i - \mC[c], c)$, where $c$ is the largest value with $\mC[c] < i$ and $S.\mselect(i,c)$ finds the $i$\nobreakdash-th occurrence of character $c$ in string $S$. If $k \ge 0$ is the smallest value for which $j = \Psi^{k}(i) \le m$, we know that $\mSA[i]$ points to offset $\abs{T_{j}} - k$ in text $T_{j}$.

We can \emph{extract} text $T_{i}$ in $\Oh(\abs{T_{i}} \cdot t_{r})$ time by using \LF\nobreakdash-mapping \cite{Burrows1994}. We start from the endmarker at $\mBWT[i]$ and extract the text backwards as $T_{i}[\abs{T_{i}} - j] = \mBWT[\LF^{j-1}(i)]$, for $1 \le j \le \abs{T_{i}}$. As $\mSA[\LF^{j}(i)]$ points to suffix $T_{i}[\abs{T_{i}}-j, \abs{T_{i}}]$, we also find the lexicographic ranks of all suffixes of text $T_{i}$ in the process.

\Section{Space-efficient \BWT{} construction}

The FM-index was introduced as a more space-efficient alternative to the suffix array. If we need the suffix array to build the FM-index, a large part of this benefit is lost, and index construction becomes the bottleneck. To overcome the bottleneck, we can use \emph{incremental construction algorithms} that build the FM-index directly. Some of them use an adjustable amount of working space on top of the FM-index, making it possible to index text collections larger than the size of the memory.

Assume that we have built the \BWT{} of text $T$, and we want to \emph{transform} the \BWT{} into that of text $cT$, where $c$ is a character \cite{Hon2007}. We find the pointer $\mSA[i]$ to the beginning of text $T$ (where $\mBWT[i] = 0$). Then we determine the lexicographic rank $j = \mrank(cT, T) = \mC[c] + \mBWT.\mrank(i, c)$ of text $cT$ among the suffixes of text $T$. Finally we \emph{replace} the endmarker at $\mBWT[i]$ with the inserted character $c$ and \emph{insert} a new endmarker between $\mBWT[j]$ and $\mBWT[j+1]$.

We can use the transformation for \BWT{} construction in several ways. We can use \emph{batch updates} and transform the \BWT{} of text $T$ into that of text $XT$, where $X$ is a string \cite{Hon2007}. We can start with the \BWT{}s of text collections $\Acoll$ and $\Bcoll$, and \emph{merge} them into the \BWT{} of collection $\Acoll \cup \Bcoll$ \cite{Siren2009}. We can also \emph{extend} multiple texts at once by inserting a new character to the beginning of each of them \cite{Bauer2013}. In all cases, we can use either \emph{static} or \emph{dynamic} \cite{Chan2007} structures for the \BWT. Dynamic representations increase the size of the \BWT{} (e.g.~by around 1.5x in \ropebwtii{} \cite{Li2014a}), while static representations require more space overhead for buffering the updates.

\begin{figure}[t!]
\includegraphics[width=\textwidth]{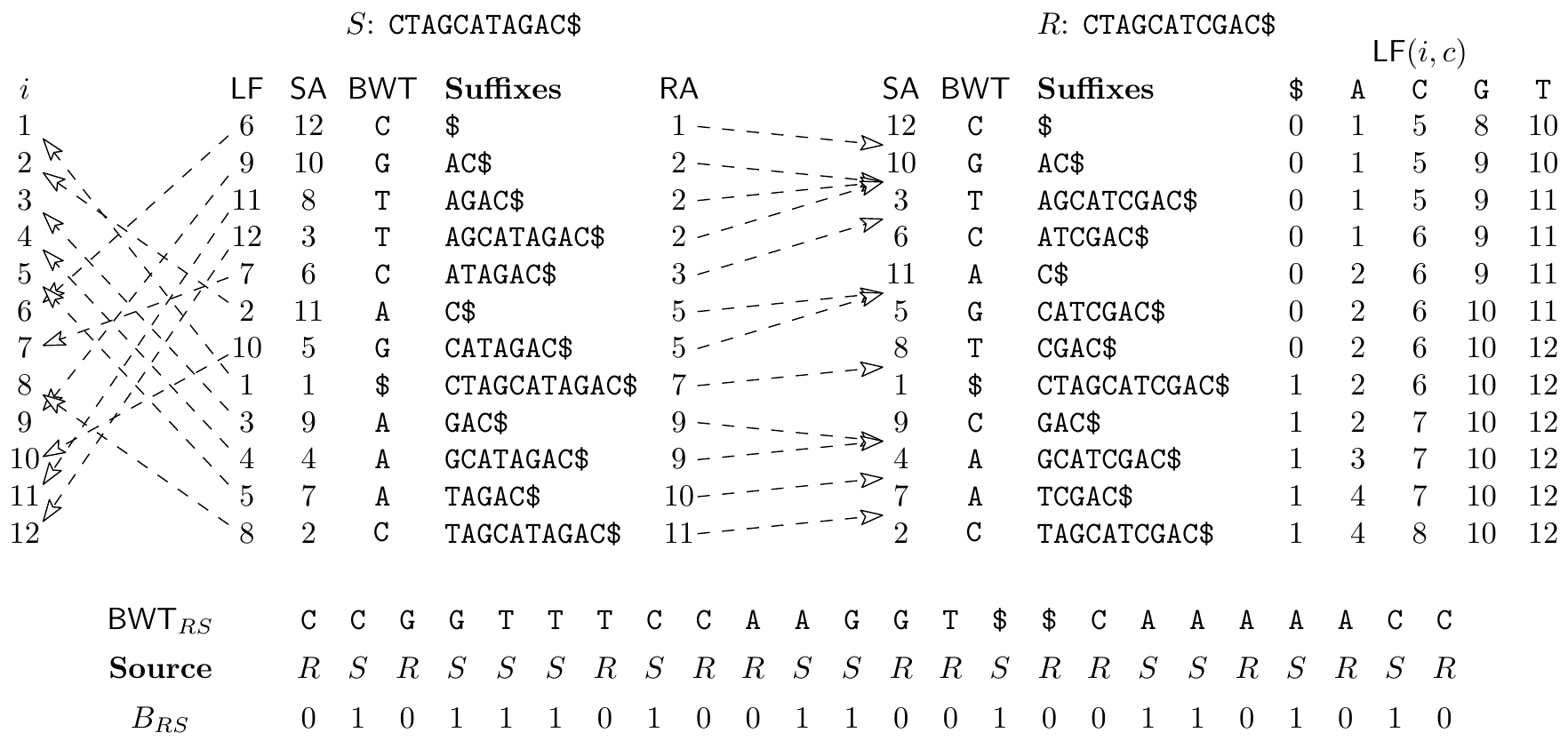}
\caption{Merging the \BWT{}s of texts $R$ and $S$. Rank array \RA{} counts the number of suffixes of $R$ that are lexicographically smaller than or equal to the given suffix of $S$. We fill it by starting with $\mRA[1] = 1$ and iterating $\mRA[\mLF_{S}(i)] = \mLF_{R}(\mRA[i], \mBWT_{S}[i])$. Interleaving bitvector $B_{RS}$ tells whether the source of a character in the merged \BWT{} is in $\mBWT_{R}$ or $\mBWT_{S}$. We build it by setting bits $i+\mRA[i]$ to 1 for all $i$.}\label{fig:merge}%
\vspace{-6pt}
\end{figure}

Assume that we want to merge the \BWT{}s of two text collections $\Acoll$ and $\Bcoll$ of total length $n_{\Acoll}$ and $n_{\Bcoll}$, respectively \cite{Siren2009}. We store the \BWT{}s in two-level arrays, where the first level contains pointers to $b$\nobreakdash-bit \emph{blocks}. If a \BWT{} takes $x$ bits, the space overhead from the array is $\frac{x}{b} \log x + \Oh(b)$ bits. This becomes $\Oh(\sqrt{x \log x})$ bits with $b = \sqrt{x \log x}$. The merging algorithm has three phases: search, sort, and merge. It uses $\Oh(n_{\Acoll} + n_{\Bcoll} t_{r})$ time and $\min(n_{\Bcoll} \log n_{\Acoll}, n_{\Acoll} + n_{\Bcoll}) + \Oh(\sqrt{x \log x})$ bits of working space in addition to the \BWT{}s and the structures required to use them as FM-indexes. See Figure~\ref{fig:merge} for an example with two texts.

\smallbreak\noindent\textbf{Search.} We search for all texts of collection $\Bcoll$ in $\mBWT_{\Acoll}$, and output the lexicographic rank $\mrank(X, \Acoll)$ for each suffix $X$ of $\Bcoll$. This takes $\Oh(n_{\Bcoll} t_{r})$ time. We either need the collection in plain form, or extract the texts from $\mBWT_{\Bcoll}$ in the same asymptotic time.

\smallbreak\noindent\textbf{Sort.} We build the \emph{rank array} (\RA) of $\Bcoll$ relative to $\Acoll$ by sorting the ranks. The rank array is defined as $\mRA_{\Bcoll \mid \Acoll}[i] = \mrank(X, \Acoll)$, where $\mSA_{\Bcoll}[i]$ points to suffix $X$. The array requires $n_{\Bcoll} \log n_{\Acoll}$ bits of space, and we can build it in $\Oh(sort(n_{\Bcoll}, n_{\Acoll}))$ time, where $sort(n, u)$ is the time required to sort $n$ integers from universe $[0,u]$. If we extracted the texts from $\mBWT_{\Bcoll}$, we can write the ranks directly into the rank array, making this phase trivial. We can also encode the rank array as a binary sequence $B_{\Acoll \cup \Bcoll}$ of length $n_{\Acoll} + n_{\Bcoll}$. This \emph{interleaving bitvector} is built by setting $B_{\Acoll \cup \Bcoll}[i + \mRA_{\Bcoll \mid \Acoll}[i]] = 1$ for $1 \le i \le n_{\Bcoll}$. If $B_{\Acoll \cup \Bcoll}[j] = 1$, we know that $\mSA_{\Acoll \cup \Bcoll}[j]$ points to a suffix of $\Bcoll$.

\smallbreak\noindent\textbf{Merge.} We interleave $\mBWT_{\Acoll}$ and $\mBWT_{\Bcoll}$ according to the rank array. If $\mRA_{\Bcoll \mid \Acoll}[i] = j$, the merged \BWT{} will have $j$ characters from $\mBWT_{\Acoll}$ before $\mBWT_{\Bcoll}[i]$. This phase takes $\Oh(n_{\Acoll} + n_{\Bcoll})$ time. By reusing the blocks of $\mBWT_{\Acoll}$ and $\mBWT_{\Bcoll}$ for $\mBWT_{\Acoll \cup \Bcoll}$, we can merge the \BWT{}s almost in-place. The total working space is $\Oh(\sqrt{x \log x})$ bits, where $x$ is the maximum of the sizes of $\mBWT_{\Acoll}$ and $\mBWT_{\Bcoll}$ in bits.

\Section{Large-scale \BWT{} merging}

If we split a text collection $\Acoll$ of total length $n$ into $p$ \emph{subcollections} of equal size, we can build $\mBWT_{\Acoll}$ incrementally by merging the \BWT{}s of the subcollections. This takes $\Oh((p+t_{r})n)$ time and uses essentially $\min(\frac{n}{p} \log n, n)$ bits of working space.

When the collection is large, the space overhead of the construction algorithm often determines whether we can build the \BWT{}.
Even if a static encoding of the \BWT{} fits in memory, a dynamic encoding may already be too large. The space overhead from the rank array or the interleaving bitvector (or their equivalents in the other space-efficient algorithms) may also be too much. We can make the rank array fit in memory by increasing the number of subcollections, but that can make the construction too slow.

We can reduce the overhead by writing the lexicographic ranks to \emph{disk}. If we sort the ranks on disk, we just need to scan the rank array once during the merge phase. We can also \emph{compress} the ranks before writing them to disk and \emph{interleave} the sorting with the search and merge phases. We now describe the key ideas for fast and space-efficient \BWT{} construction.

\smallbreak\noindent\textbf{Search.} Instead of searching for every text in collection $\Bcoll$ separately, we can search for the \emph{reverse trie} of the collection. Assume that there are $m_{\Acoll}$ texts in collection $\Acoll$ and $m_{\Bcoll}$ texts in collection $\Bcoll$. The \emph{root} of the trie corresponds to suffix $\$$, which has lexicographic rank $m_{\Acoll}$ in $\Acoll$ and corresponds to lexicographic range $[1,m_{\Bcoll}]$ in $\Bcoll$.

Assume that we have a \emph{node} of the trie corresponding to suffix $X$, lexicographic rank $r$, and lexicographic range $[sp,ep]$. As suffix $X$ occurs $ep+1-sp$ times in collection $\Bcoll$, we can output a \emph{run} of ranks $(r, ep+1-sp)$. Afterwards, we proceed to the \emph{children} of the node. For each character $c \in \Sigma$, we create a node corresponding to suffix $cX$, rank $\mLF_{\Acoll}(r,c)$, and range $[\mLF_{\Bcoll}(sp-1, c) + 1, \mLF_{\Bcoll}(ep, c)]$. Searching the branches of the trie can be done in parallel using multiple \emph{threads}.

\smallbreak\noindent\textbf{Buffering.} To reduce disk I/O and space usage, we buffer and compress the lexicographic ranks before writing them to disk. Each thread has two buffers: a \emph{run buffer} and a \emph{thread buffer}. The run buffer stores the runs as pairs of integers $(r, \ell)$. Once the run buffer becomes full, we sort the runs by \emph{run heads} $r$, use \emph{differential encoding} for the run heads, and encode the differences and run lengths with a \emph{prefix-free code}. The compressed run buffer is then merged with the similarly compressed thread buffer.

Once the thread buffer becomes full, we merge it with the global \emph{merge buffers}. There are $k$ merge buffers $M_{1}$ to $M_{k}$, with buffer $M_{i}$ containing $2^{i-1}$ thread buffers. The merging starts from $M_{1}$. If $M_{i}$ is empty, the thread swaps its thread buffer with the empty buffer and returns to the search phase. Otherwise it merges $M_{i}$ with its thread buffer, clearing $M_{i}$, and proceeds to $M_{i+1}$. If the a thread reaches $M_{k+1}$, it writes its thread buffer to disk and returns back to work.

\smallbreak\noindent\textbf{Merge.} The ranks are stored in sorted order in multiple files on disk. For interleaving the \BWT{}s, we need to merge the files and to scan through the rank array. We can also use multiple threads here. One thread reads the files and performs a \emph{multiway merge} using a priority queue, producing a stream of lexicographic ranks. Another thread consumes the stream and uses it to interleave the \BWT{}s. If the disk is fast enough, we may want to use multiple threads for the multiway merge.

\Section{Implementation}

We have implemented the improved merging algorithm as a tool for merging the \BWT{}s of large read collections. The tool, \BWTmerge{}, is written in C++, and the source code is available on GitHub.\footnote{\url{https://github.com/jltsiren/bwt-merge}} The implementation uses the \emph{SDSL library} \cite{Gog2014b} and the new features in C++11. As a result, it needs a fairly recent C++ compiler to compile. We have successfully built \BWTmerge{} on Linux and OS~X using g++.

The target environment of \BWTmerge{} is a \emph{single node} of a \emph{computer cluster}. The system should have tens of CPU cores, hundreds of gigabytes of memory, and hundreds of gigabytes of local disk space for temporary files. The number of search threads is equal to the number of CPU cores, while the merge phase uses just one producer thread and one consumer thread. \BWTmerge{} can be adapted to many other environments by adjusting the number and the size of the buffers.

The internal alphabet of \BWTmerge{} is \texttt{012345}, which corresponds to either \texttt{\$ACGTN} or \texttt{\$ACGNT}, depending on where the \BWT{}s come from. \BWT{}s using different alphabetic orders cannot be merged. We use simple byte-level codes for \emph{run-length encoding} the \BWT{}s. The encoding of run $(c, \ell)$, where $c$ is the character value and $\ell$ is the length, depends on the length of the run. If $\ell \le 41$, the run is encoded in a single byte as $6 \cdot (\ell-1) + c$. Longer runs start with byte $6 \cdot 41 + c$, followed by the encoding of $\ell-42$. The remaining run length is encoded as a sequence of bytes, with the low 7 bits containing data and the high bit telling whether the encoding continues in the next byte. The compressed buffers use the same 7+1\nobreakdash-bit code for both the differentially encoded run heads and the run lengths.

For \rank/\select{} support, we divide the \BWT{}s into 64\nobreakdash-byte blocks of compressed data, ensuring that the runs do not cross block boundaries. For each block $i$, we store the total number of characters in blocks $1$ to $i-1$ as $n_{i}$, as well as the cumulative character counts $c_{i} = \mBWT.\mrank(n_{i},c)$ for $0 \le c \le 5$. These increasing sequences are stored using the \emph{sdarray} encoding \cite{Okanohara2007}. To compute $\mBWT.\mrank(j,c)$, we start with a \rank{} query on the $n_{i}$ sequence to find the block. A \select{} query on the same sequence transforms $j$ into a block offset, while a \select{} query on the $c_{i}$ sequence gives the rank up to the beginning of the block. We then decompress the block to answer the query. \select{} queries and accessing the \BWT{} work in a similar way. There are also optimizations for e.g.~computing $\mrank(i,c)$ for all characters $c$, and for finding the children of a reverse trie node corresponding to a short lexicographic range.

We use two-level arrays with 8\nobreakdash-megabyte blocks to store the \BWT{}s and the compressed buffers, managing the blocks using \texttt{mmap()} and \texttt{munmap()}. This reduces the space overhead by tens of gigabytes over using \texttt{malloc()} and \texttt{free()}.
When multiple threads allocate memory in small enough blocks, the multithreaded \emph{glibc} implementation of \texttt{malloc()} creates a number of additional heaps that will never grow larger than 64~MB. Each thread tries to reuse the heap it used for the last allocation. If a heap is full or the thread cannot acquire the mutex, it moves to the next heap. With our workload of tens of threads allocating and freeing hundreds of gigabytes of memory, this created thousands of heaps with holes in the middle. Most \texttt{free()} calls did not release the memory back to the operating system, while it took a while before any thread could reuse the holes created by the \texttt{free()} calls.

\Section{Experiments}

We used a system with two 16\nobreakdash-core AMD Opteron 6378 processors and 256 gigabytes of memory. The system was running Ubuntu 12.04 on Linux kernel 3.2.0. We used a development version of \BWTmerge{}  equivalent to v0.3, and the versions of the other tools that were available on GitHub in October 2015. All software was compiled with gcc/g++ version 4.9.2. We stored the input/output files on a distributed Lustre file system and used a local 0.5~TB disk for temporary files. \BWTmerge{} used 32 threads, while the other \BWT{} construction tools were limited by design to 4 or 5 threads.
The merging times include verification by querying the \BWT{}s with 2~million $32$\nobreakdash-mers.

\begin{table}[t!]
\begin{center}
\caption{Datasets. The amount of sequence data, the number of reads, and the size of the \BWT{} in the native format and in the Read Server format. RLO indicates that the reads are sorted in reverse lexicographic order. The numbers in parentheses are estimates.}\label{table:datasets}\smallskip%
{
\renewcommand{\baselinestretch}{1}\footnotesize
\begin{tabular}{c|cc|cc|cc}
\hline
 & \multicolumn{2}{c|}{\textbf{Data}} & \multicolumn{2}{c|}{\textbf{Native \BWT}} &
 \multicolumn{2}{c}{\textbf{Read Server}} \\
\textbf{Dataset} & \textbf{Size} & \textbf{Reads} & \textbf{Unsorted} & \textbf{RLO} & \textbf{\BWT} & \textbf{\FMI} \\
\hline
\CEU: All           &  771 Gbp & 7.63G &  136 GB & 65.9 GB &      -- &       -- \\
NA12878             &  284 Gbp & 2.81G & 50.3 GB & 25.5 GB &      -- &       -- \\
NA12891             &  242 Gbp & 2.40G & 42.4 GB & 19.7 GB &      -- &       -- \\
NA12892             &  245 Gbp & 2.42G & 43.8 GB & 20.7 GB &      -- &       -- \\
Merged              &  771 Gbp & 7.63G &  129 GB & 58.9 GB &      -- &       -- \\
\hline
\RS: AA, TT, AT, TA & 1.49 Tbp & 16.2G &      -- &  136 GB &  140 GB &   170 GB \\
AA                  &  433 Gbp & 4.69G &      -- & 38.5 GB & 39.9 GB &  48.3 GB \\
TT                  &  432 Gbp & 4.68G &      -- & 38.7 GB & 40.0 GB &  48.4 GB \\
AT                  &  275 Gbp & 2.98G &      -- & 26.6 GB & 26.9 GB &  32.6 GB \\
TA                  &  355 Gbp & 3.84G &      -- & 32.7 GB & 33.5 GB &  40.6 GB \\
Merged              & 1.49 Tbp & 16.2G &      -- &  117 GB &  126 GB & (152 GB) \\
\hline
\RS: *A, *C         & 2.45 Tbp & 26.5G &      -- &  225 GB &  232 GB &   281 GB \\
Merged              & 2.45 Tbp & 26.5G &      -- &  181 GB &  197 GB & (239 GB) \\
\hline
\RS: *G, *T         & 2.44 Tbp & 26.5G &      -- &  226 GB &  232 GB &   281 GB \\
Merged              & 2.44 Tbp & 26.5G &      -- &  180 GB &  197 GB & (238 GB) \\
\hline
\end{tabular}}
\end{center}%
\vspace{-16pt}
\end{table}

Our datasets come from phase 3 of the \emph{1000 Genomes Project} \cite{1000GP2015}. \CEU{} contains 101~bp reads from high-coverage sequencing of the \emph{CEU trio} (individuals NA12878, NA12891, and NA12892). We downloaded the gzipped FASTQ files (run accessions SRR622457, SRR622458, and SRR622459). For each individual, we concatenated the files and corrected the sequencing errors with BFC \cite{Li2015} (\texttt{bfc -s 3g -t 16}). \RS{} is from the \emph{Read Server} project, which uses all low-coverage and exome data from the phase 3. There are 53.0~billion unique reads for a total of 4.88~Tbp. The reads are in 16 run-length encoded \BWT{}s built by using the \emph{String Graph Assembler} (SGA) \cite{Simpson2012}, partitioned by the last two bases. See Table~\ref{table:datasets} for further details on the datasets.

\smallbreak\noindent\textbf{Parameters.} For testing different parameter values, we took four \BWT{} files (AA, TT, AT, and TA) containing a total of 1.49~Tbp from the \RS{} dataset, and converted them to the \emph{native format} of \BWTmerge. This format includes the \rank/\select{} structures required by the FM-index. We then merged the \BWT{}s (in the given order). We used 128~MB or 256~MB run buffers and 256~MB or 512~MB thread buffers. The number of merge buffers was chosen so that the files on disk were always merged from either 8~GB or 16~GB of thread buffers.

\begin{figure}[t!]
\begin{center}
\includegraphics{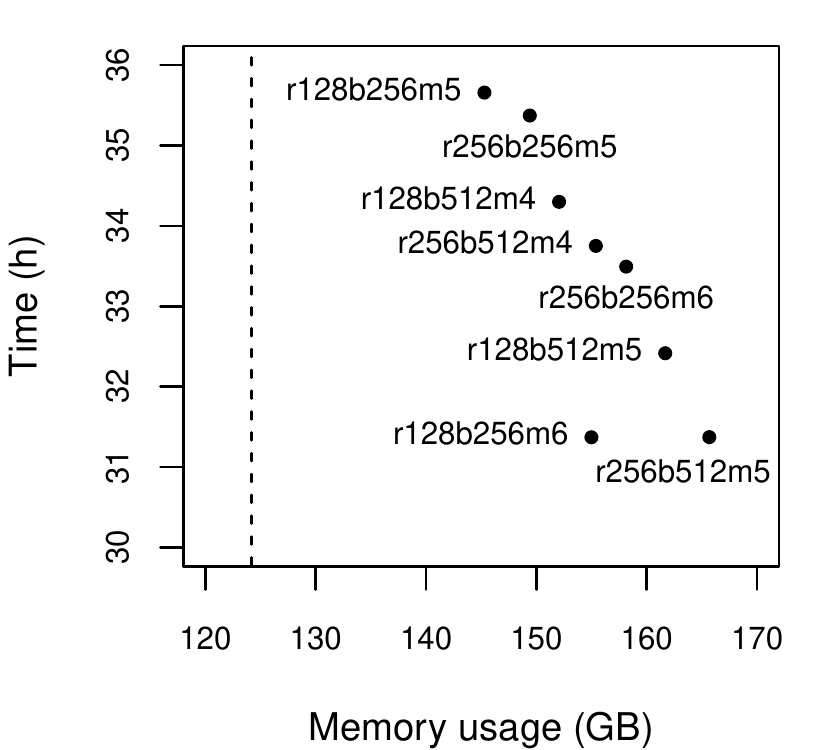}%
\hspace{-0.3in}%
\includegraphics{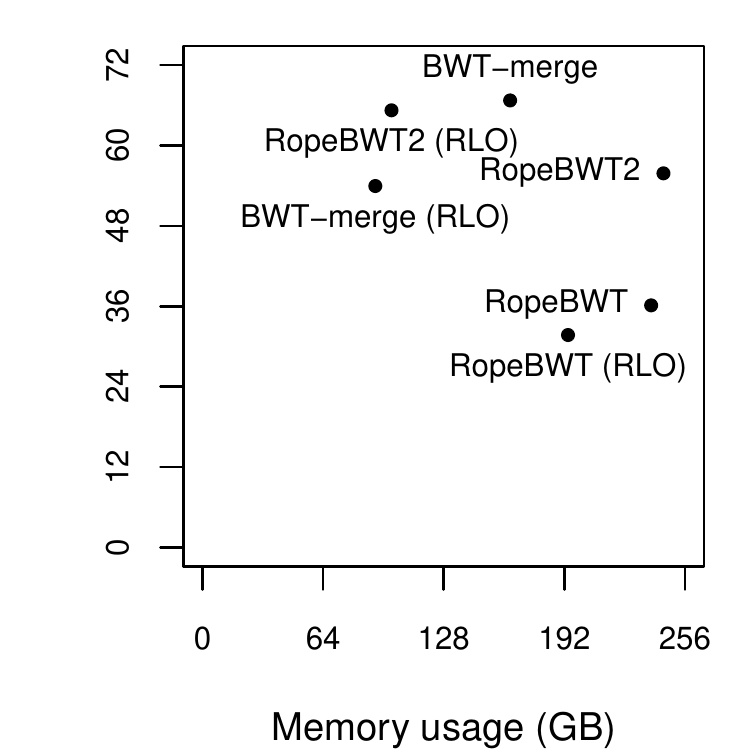}
\end{center}%
\vspace{-12pt}
\caption{Time/space trade-offs. Left: Merging four \BWT{} files (1.49~Tbp). Label rXbYmZ denotes $X$~MB run buffers, $Y$~MB threads buffers, and $Z$ merge buffers. The dashed line marks the total size of the last pair of \BWT{}s to be merged. Right: Building the \BWT{} of the 771~Gbp \CEU{} dataset. RLO indicates reverse lexicographic order.}\label{fig:benchmark}%
\vspace{-6pt}
\end{figure}

The results can be seen in Figure~\ref{fig:benchmark} (left). The average speed for inserting 1.06~Tbp into file AA ranged from 8.27~Mbp/s to 9.40~Mbp/s, depending on the parameter values. Memory overhead was 21.1~GB to 41.5~GB on top of the 124.2~GB required by the last pair of \BWT{}s.
The temporary files used 287 to 306~gigabytes of disk space. Thread buffer size and the number of merge buffers were the most important parameters. The larger the individual files are, the more memory the search phase uses and the faster the merge phase is. Increasing run buffer size to 256~MB made the search phase faster with 512~MB thread buffers and slower with 256~MB thread buffers.
For the further experiments, we chose 128~MB run buffers, 256~MB thread buffers, and 6 merge buffers (overhead 30.8~GB).

\smallbreak\noindent\textbf{Comparison.} In the next experiment, we compared \BWTmerge{} to the fastest \BWT{} construction tools on general hardware \cite{Li2014a}. We built the \BWT{} of the \CEU{} dataset using \ropebwt{} \cite{Li2011-2013} with parameters \texttt{-btORf -abcr} and \ropebwtii{} \cite{Li2014a} with parameters \texttt{-bRm10g}. We also built individual \BWT{}s using \ropebwt{} and merged them with \BWTmerge. All tools were set to write the \BWT{}s in their preferred formats.

The results are in Figure~\ref{fig:benchmark} (right). When the reads are in the original order, \BWTmerge{} is 1.85x slower and 1.46x more space-efficient than \ropebwt. \ropebwt2{} ran out of memory just before finishing. It would have been about 1.2x faster and 1.5x less space-efficient than \BWTmerge. The running time of \BWTmerge{} was split evenly between \BWT{} construction and merging.

When \ropebwt{} and \ropebwt2{} sort the reads in \emph{reverse lexicographic order} (RLO) to improve compression, all tools improve their performance. \BWTmerge{} becomes 1.70x slower and 2.12x more space-efficient than \ropebwt{}, and 1.21x \emph{faster} and 1.09x more space-efficient than \ropebwt2.
Again, \BWTmerge{} spent around half of the time building the individual \BWT{}s and another half merging them. Note that \BWTmerge{} builds a \BWT{} for the concatenation of three input files that are in RLO, while the \BWT{}s produced by the other tools are completely in RLO. Maintaining RLO during merging would reduce the size of the final \BWT{} from 58.9~GB to 54.4~GB.

\smallbreak\noindent\textbf{Read Server.} In the last experiment, we merged the 16 \BWT{} files in the \RS{} dataset into two files (AA, CA, TA, GA, AC, CC, GC, and TC into the first file; TT, GT, CT, AT, TG, GG, CG, AG into the second one). Merging the \BWT{}s took 81.3~hours and 83.0~hours, required 221~GB and 219~GB of memory, and used 297~GB and 300~GB of disk space, respectively. This reduced the size of the FM-indexes from around 560~GB to 480~GB. By converting the \BWT{}s to the native format of \BWTmerge{}, we further reduced the size of the indexes to 360~GB.
This makes it possible to host the indexes on two servers instead of the original three.

\Section{Conclusions}

We have proposed an improved \BWT{} merging algorithm for large read collections. Our implementation of the algorithm in the \BWTmerge{} tool is fast enough to be used with terabases of sequence data. It requires only 30~gigabytes of memory on top of the \BWT{}s to be merged. As \BWT-based indexes access large arrays in a random fashion, they must reside in memory in most applications. Hence \BWTmerge{} can build the index on the same system as it is going to be used.

\BWTmerge{} can be used as a part of a \BWT{} construction algorithm. We split the read collection into subcollections, build the \BWT{}s of the subcollections, and merge the results. The resulting algorithm is typically slower but more space-efficient than the existing algorithms.

The most important feature of our algorithm is its low memory usage. With it, we can build the \BWT{}s of much larger read collections than before on commonly available hardware. As a concrete example, we merged the 16 Read Server \BWT{} files into two files. This reduced the number of servers required to host the indexes from three to two, and also improved the query performance of the servers.

In the future, we are going to extend \BWTmerge{} to support different \emph{text orders}, and to optionally \emph{remove duplicate texts} from the merged collection. The current algorithm maintains the existing order by inserting the texts from $\mBWT_{\Bcoll}$ after the texts in $\mBWT_{\Acoll}$. This makes it easy to determine the original text identifiers without having to store a permutation. Other text orders are useful for different purposes.

Read Server stores the reads in \emph{reverse lexicographic order} to improve compression \cite{Cox2012}. We can maintain this order with a few changes to the search phase \cite{Li2014a}. Sorting the reads by their \emph{reverse complements} also improves compression in a similar way. In this order, $\mSA[\mBWT.\mselect(i,0)]$ points to the beginning of the reverse complement of read $T_{i}$, if the collection includes the reverse complement of every read \cite{Li2014a}. With \emph{lexicographic order}, we can determine $\mSA[i]$ without samples by using $\mLF$ instead of $\Psi$, which is often faster in practice.

We can also sort the reads by their likely positions in a reference genome. This \emph{position order} is useful for both compression and storing the \emph{pairing information} for the reads. Consider a graph with reads as its nodes and edges between paired reads. If we sort the reads in position order, most edges will be close to the diagonal of the edge matrix. Such matrices are very easy to compress.

\newpage
\Section{References}
\bibliographystyle{IEEEtran}
\bibliography{paper}

\end{document}